\documentclass[twocolumn,aps,superscriptaddress,pra,showpacs]{revtex4}

\usepackage{epsfig}
\usepackage{color}

\begin{document}

\title{Visibility bound caused by a distinguishable noise particle}


\author{Miroslav Gavenda}
\author{Lucie \v{C}elechovsk\'{a}}
\affiliation{Department of Optics, Faculty of Science,
Palack\' y University, 17.~listopadu 12, 771~46 Olomouc, Czech Republic}
\author{Jan Soubusta}
\affiliation{Joint Laboratory of Optics of Palack\'{y} University and
Institute of Physics of Academy of Sciences of the Czech Republic,
17. listopadu 50A, 779 07 Olomouc, Czech Republic}
\author{Miloslav Du\v{s}ek}
\affiliation{Department of Optics, Faculty of Science,
Palack\' y University, 17.~listopadu 12, 771~46 Olomouc, Czech Republic}
\author{Radim Filip}
\affiliation{Department of Optics, Faculty of Science,
Palack\' y University, 17.~listopadu 12, 771~46 Olomouc, Czech Republic}

\date{\today}

\begin{abstract}
We investigate how distinguishability of a ``noise'' particle
degrades interference of the ``signal'' particle. The signal,
represented by an equatorial state of a photonic qubit, is mixed
with noise, represented by another photonic qubit, via linear
coupling on the beam splitter.  We report on the degradation of
the ``signal'' photon interference depending on the degree of
indistinguishability between ``signal'' and ``noise'' photon. 
When the photons are principally completely
distinguishable but
technically indistinguishable the visibility drops to the value
$1/\sqrt{2}$.   
As the photons become more indistinguishable the maximal visibility 
increases and reaches the unit value for 
completely indistinguishable photons.
We have examined this effect
experimentally using setup with fiber optics two-photon
Mach-Zehnder interferometer.
\end{abstract}  \maketitle

\section{Introduction}

The key property of the quantum world is the existence of
superpositions of states. This property plays a crucial role in
quantum information transfer and processing. However, many systems
cannot preserve coherent superpositions for long time  due to
decoherence \cite{Zurek_03,Schlosshauer_04}. This process of
decoherence may have several reasons: It can arise from
fluctuations of external  macroscopic physical parameters of the
system (dephasing) \cite{Wooters_79} or it can appear as the
result of a  coherent coupling between the system and the
environment \cite{Zurek_82,Joos_85} (experimentally tested in
Ref.~\cite{Brune_96}) or it can be evoked by mixing the quantum
system with some other system representing noise. In this paper we
will focus on the latter.  Contrary to the standard model of
decoherence our mechanism is not based on coherent interaction between the
system and environment \cite{Zurek_82, Schlosshauer_04}. 
No spontaneous emission from the system
\cite{Pfau_94} or collision to the environment appear
\cite{Chapman_95}. The only relevant source of
decoherence is the added noise and its distinguishability from the signal \cite{Sciarrino_09}.
The problem of noise is of a special interest for quantum
communication \cite{comp}.
The physical scenario we describe is relevant for boson like particles, to
extend it for fermions their specific properties had to be taken into
account
\cite{Banuls_07}.

\section{Theoretical framework}
The simplest system in quantum information is a qubit. In many
applications and experiments (like in quantum key distribution,
tests of Bell inequalities, etc.) only a subset of all possible
states of a qubit is used. Namely, the set of equatorial states
(they lie on the equator of the Bloch sphere) represented by a
coherent superposition
$[|0\rangle+\exp(i\varphi)\,|1\rangle]/\sqrt{2}$ of two basis
states $|0\rangle$ and $|1\rangle$. Phase $\varphi$ is used to
encode information.
We investigate how noise can affect the coherence of equatorial
state of qubit. We suppose no fluctuation of phase and no
interaction with the environment (therefore no entanglement can
appear between the qubit and the environment). Let us represent
our qubit by a single particle distributed between two modes $A$
and $B$. Its equatorial state can be described as
$|\Psi\rangle_{AB}=[|1,0\rangle_{AB}+\exp(i\varphi)\,|0,1\rangle_{AB}]/\sqrt{2}$, 
where $0$ and $1$ represent the number of particles.
The noise is caused by another particle in mode $B'$ which can be
confused with the original particle. We suppose that modes $B$ and
$B'$ are {\em principally} distinguishable but {\em technically}
indistinguishable. It means our detectors cannot discriminate them.
We will consider the situation when a ``noise'' particle is
created in mode $B'$ and subsequently one particle is annihilated
either from mode $B$ or $B'$ (our device cannot distinguish
between them).

To examine this situation experimentally, we use a Mach-Zehnder
(MZ) interferometer and a source of photon pairs
(Fig.~\ref{scheme}). The ``noise'' photon with variable
distinguishability from the ``signal'' one is fed into one arm of
the interferometer. If the ``noise'' particle was created in mode
$B$ it would be {\em principally indistinguishable} from the
``signal'' particle in mode $B$. For a bosonic field the total
quantum state after the creation process
$a^{\dagger}_B|\Psi\rangle_{AB}$ reads
$|\Psi'\rangle_{AB}=[|1,1\rangle_{AB}+\sqrt{2}\exp(i\varphi)\,|0,2\rangle_{AB}]/\sqrt{3}$.
It has full coherence: the phase information is fully preserved.
The indistinguishable photon can be eliminated by the act of
annihilation $a_B|\Psi'\rangle_{AB}$ ending up with state
$|\Psi''\rangle_{AB}=[|1,0\rangle_{AB}+2\exp(i\varphi)\,|0,1\rangle_{AB}]/\sqrt{5}$.
It can be further probabilistically converted to the original
state of the signal qubit applying attenuation $\eta_B=1/4$ in
mode $B$. Since the both particles are indistinguishable, it does
not matter which one has been actually taken out. The visibility
of interference can reach unity again. We define the single photon
visibility by the standard formula $V =
(P_{\max}-P_{\min})/(P_{\max}-P_{\min})$ where
$P_{\max}=\max_{\varphi}P(\varphi)$,
$P_{\min}=\min_{\varphi}P(\varphi)$ with
 $P(\varphi)$ being the probability to detect photon at detector $D$
 depending on phase $\varphi$.

If the ``noise'' particle  is {\em principally distinguishable}
from the ``signal'' one, it can be described by creation operation
$a^{\dagger}_{B'}|\Psi\rangle_{AB}|0\rangle_{B'}$. In principle,
it could be filtered out, because it differs in its properties
from the ``signal'' particle. But, quite typically, our filters
are not selective enough to enable it. If the disturbing particle
is only {\em technically indistinguishable}, the total state of
the system is transformed to
$|\Phi'\rangle=[|1,0,1\rangle_{ABB'}+\exp(i\varphi)\,|0,1,1\rangle_{ABB'})]/\sqrt{2}$
after the creation in mode $B'$. Because we are not able to
discriminate modes $B$ and $B'$, to remove a single particle we
just randomly annihilate a single particle either from $B$ or $B'$
(with no prior knowledge this strategy is fully symmetrical).
Further, without any access to mode $A$, this process can be
described by two ``subtraction'' operators: $S_1=a_B\otimes
1_{B'}$ and $S_2=1_{B}\otimes a_{B'}$ acting with equal
probabilities. Applying these operators on
$\rho'=|\Phi'\rangle\langle\Phi'|$, i.e. $S_1\rho'
S_1^{\dagger}+S_2\rho' S_2^{\dagger}$, one gets the resulting
mixed state
$\rho''=2/3\,|\Psi\rangle_{AB}\langle\Psi|\otimes|0\rangle_{B'}\langle
0|+ 1/3\,|00\rangle_{AB}\langle 00|\otimes
|1\rangle_{B'}\langle 1|$. Because our detectors cannot
distinguish whether the particle came from mode $B$ or $B'$, the
visibility of interference is now $V''=2/3$. It can be
probabilistically enhanced by a proper attenuation
$\eta_B=\eta_{B'}$ in modes $B$ and $B'$. This transforms the
total state to
\begin{eqnarray}\label{st1}
\rho''' &=& \frac{1}{2\eta_B}
\left[|100\rangle\langle 100|+\eta_B\,|010\rangle\langle 010|+
    \eta_B\,|001\rangle\langle 001|\right. \nonumber\\
& & + \left.\sqrt{\eta_B} \left(e^{i\varphi}\,|010\rangle\langle 100|+
e^{-i\varphi}\,|100\rangle\langle 010|\right)\right].
\end{eqnarray}
Clearly, if $\eta_B=1/2$ one balances the probability of having a
particle in mode $A$ with the probability of having it either in
mode $B$ or $B'$.
Then the visibility is maximal and reaches the
value
\begin{equation}\label{inc}
V_{dis}=\frac{1}{\sqrt{2}}.
\end{equation}

In comparison to the previous case of indistinguishable particles,
this reduction of visibility represents a fundamental impact of
the principal distinguishability of the ``signal'' and ``noise''
particles. The loss of coherence is a result of an elementary
ignorance of our measurement apparatus.
There is a difference between our result and the one reported in
\cite{Zou_1991}, where visibility completely vanishes for distinguishable 
photons. 
According to Englert's inequality $V^2+K^2\leq 1$ \cite{Englert_1996,Scully_1991}, this elementary
visibility reduction corresponds to overall which-way knowledge
$K<1/\sqrt{2}$ accessible in the experiment. 

If $N$ completely distinguishable particles are created
simultaneously in modes associate to mode $B$ and, subsequently,
$N$ particles are simultaneously annihilated in these modes and
mode $B$, visibility rapidly decreases with increasing $N$:
$V_{dis}(N)=\frac{1}{\sqrt{N+1}}$. On the other hand if
the particles are created and annihilated subsequently, i.e. after
any single-particle creation a single particle is always
annihilated, visibility is decreasing even faster:
$V_{dis}(N)=\left(\frac{1}{\sqrt{2}}\right)^N$.

\section{Experimental implementation} 
\begin{figure}
\centerline{\psfig{width=6cm,angle=0,file=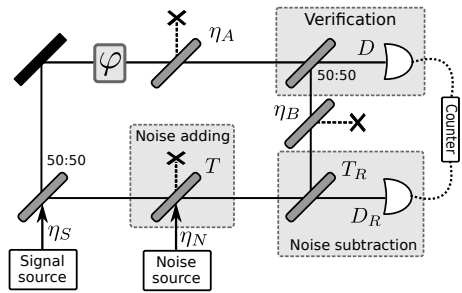}}
\caption{The simple scheme of Mach-Zehnder two-photon interferometer
illustrates the process of interference measurement.}
\label{scheme}
\end{figure}

In the experiment, mode $A$ is represented by the upper arm of the
interferometer (Fig.~\ref{scheme}) and modes $B, B'$ by the lower
arm (they are distinguishable in time domain). Creation and
annihilation process is emulated by a beam splitter. The action of
the beam splitter can be described by a unitary operator
$U=\exp[\theta(a^{\dagger} a_{\mathrm{aux}} -
a_{\mathrm{aux}}^{\dagger} a)]$, where ``aux'' denotes the
auxiliary mode and $\theta$ is related to the intensity
transmittance by the formula $T=\cos^2 \theta$. If $|\theta| \ll
1$ and there is a proper state in the auxiliary mode, $U$ well
approximates action of creation, $a^{\dagger}$, or annihilation,
$a$, operators. The coincidence measurement guaranties that only
those situations are taken into account, when exactly one photon
is annihilated and one photon is detected at the output of the
interferometer.

To be realistic and comparable with experimental results, the
theoretical prediction must take into account a finite coupling
(i.e. transmittances and reflectances of beam splitters) as well
as all insertion losses.
In  Fig.~\ref{scheme} the signal source generates single photons
which are coupled, with efficiency $\eta_{S}$, to the
interferometer and afterwards they are split equally likely to the
upper or lower arm of the interferometer. In the upper arm we can
set the phase shift $\varphi$ and adjust losses by a beam splitter
with transmissivity $\eta_{A}$ (to achieve maximum interference).
The noise source feeds single photons into the lower arm of the
interferometer with coupling efficiency $\eta_{N}$ and then the
photons are coupled by a beam splitter with transmissivity $T$ to
the signal photons. The internal losses of the interferometer are
modelled by a beam splitter with transmissivity $\eta_{B}$. As was
indicated in the introduction, in order to subtract noise we
suggested to annihilate one photon from the lower arm of the
interferometer. This is accomplished by inserting another beam
splitter which transmits photons with ratio $T_{R}$ to another
detector $D_{R}$. To evaluate the effect of the noise subtraction
we measure visibility of the signal from detector $D$ conditioned
on the detection event from the detector $D_{R}$. Calculation for
fully indistinguishable ``noise'' photon leads to visibility
\begin{equation}
V_{{ind}} =
\frac{4\sqrt{\eta_{A}\eta_{B}T(1
-T_{R})}}{\eta_{A}+4\eta_{B}T(1-T_{R})}.
\end{equation} 
Optimizing the values of free parameters 
we can reach 
\begin{equation}
V^{max}_{{ind}}=1 
\end{equation}
for
 $\eta_{A}=\eta_{B}T$, $T_{R}=3/4$. 
The perfect visibility is achieved, as was predicted
also in the previous discussion of simplified model. In the fully
distinguishable scenario the visibility reads 
\begin{equation}
V_{{dis}} =
\frac{2\sqrt{\eta_{A} \eta_{B}T(1-T_{R})}}
{\eta_{A}+2\eta_{B}T(1-T_{R})}. 
\end{equation}
If we optimize the values of free
parameters we
can reach 
\begin{equation}
V^{max}_{{dis}}=\frac{1}{\sqrt{2}}
\end{equation}
for
$\eta_{A}=\eta_{B}T$, $T_{R}=1/2$.  We can
see the drop in visibility to the value $1/\sqrt{2}$. 

In practice,
the photons are partially indistinguishable. 
We can describe this situation as a mixture of the two limit cases: With
probability $p$ the ``signal'' and ``noise'' photons are principally
indistinguishable, otherwise they are principally distinguishable!
We have repeated the calculation of
visibility (similarly to the previous extreme cases) for the above
defined mixture of an indistinguishable and distinguishable
``noise'' photon. The visibility reads
\begin{equation}
V(p)=\frac{2(1+p)\sqrt{\eta_{A}\eta_{B}T(1-T_{R})}}{\eta_{A}+2(1+p)\eta_{B}T(1-T_{R})}.
\end{equation}
Optimizing  the values of free parameters the visibility
reaches its maximum
\begin{equation}\label{main}
V^{max}(p)=\sqrt{\frac{1+p}{2}},
\end{equation}
 for
$\eta_{A}=\eta_{B}T$, $T_{R}=(1+2p)/[2(1+p)]$. 
The more distinguishable is the noise
photon from the signal photon the lower visibility we can obtain.
The transmissivity $T$, determining the strength of the coupling
between the noise and signal photon, has no influence on the
visibility.
\begin{figure}
  \begin{center}
    \smallskip
     \resizebox{7cm}{!}{\includegraphics*{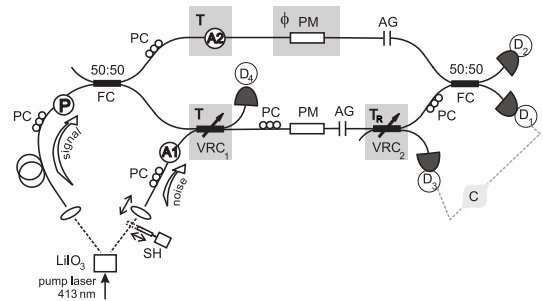}}
    \smallskip
  \end{center}
  \caption{Experimental setup. Shutter (SH), polarization controller (PC),
  polarizer (P), attenuator (A), phase modulator (PM), adjustable air-gap (AG),
  fiber coupler (FC), variable-ratio coupler (VRC), detector (D).}
  \label{setup}
\end{figure}
We have used a setup depicted in Fig.~\ref{setup} to
experimentally test the theoretically predicted visibility (\ref{main}) 
for the two extreme cases of distinguishability. 
The key part of the setup is the MZ interferometer build of 
fiber optics that allows us to simply control transmissivities $T$
and $T_{R}$ via variable-ratio couplers (VRCs) within the range 0-100$\%$.
Signal and noise photons are created by type-I degenerate
spontaneous parametric down-conversion in a nonlinear crystal of
LiIO$_{3}$ pumped by a cw Kr-ion laser (413~nm). Photons from each
pair are tightly time correlated, have the same polarization and the same 
spectrum centered at 826~nm. The degree of distinguishability of signal and
noise photons can be tuned changing the time delay between their wave-packets
at VRC$_1$. This is realized moving  a motorized translation stage connected 
to the fiber coupling system at the ``noise'' side.
All other characteristics of the photons are identical. 

Before the measurement the source of photon pairs is adjusted by
optimizing the visibility of two-photon interference at VRC$_1$ 
with splitting ratio 50:50. The visibility of Hong-Ou-Mandel (HOM) dip \cite{HOM}
reaches typically values about 98$\%$. 
Then the equality of intensities of signal and noise coupled to the fibers 
is verified measuring the count rates at detectors D$_3$ and D$_4$. 
The count rates of the noise photons at these detectors have to be
double in comparison with the count rates of the signal photons.
The required signal to noise ratio is then set tuning the intensity
transmissivity $T$ of VRC$_1$.
%
According to the theoretical proposal, the transmissivity of the upper arm 
of the MZ interferometer is also set to the value $T$ using attenuator A2. 
At this point we unbalance the interferometer setting the optimal
transmissivity $T_{R}$ of VRC$_2$. This variable ratio coupler separates a part 
of the light from the lower interferometer arm for a post-selection measurement
on the detector D$_3$. It should be stressed that these additional losses are not
compensated in the upper arm of the MZ interferometer.\\
Technical remarks: \\
(i) To accomplish proposed experiment, only one phase modulator in
the upper arm of the MZ interferometer is needed. The second phase modulator 
in the lower arm just guarantees the same dispersion in both 
interferometer arms. This trick allowed to increase the visibility approximately 
by 13~$\%$ to 94$\%$.\\
(ii) All used detectors are Perkin-Elmer single-photon counting
modules. To implement the post-selection measurement the signals from detectors 
are processed by coincidence electronics with a coincidence window 
of 2~ns.\\
(iii) The absolute phase in optical fibers is influenced by temperature
changes. Resulting undesirable phase drift is reduced by a thermal isolation 
of MZ interferometer and the residual phase drift is compensated  
by an active stabilization.
\begin{figure}
  \begin{center}
    \smallskip
     \resizebox{7cm}{!}{\includegraphics*{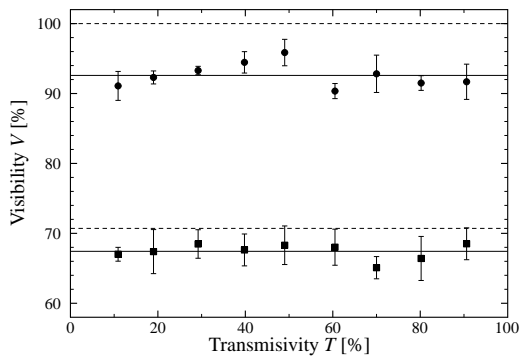}}
    \smallskip
  \end{center}
  \caption{Visibility $V$ as a function of the transmissivity $T$.
  Symbols denote experimental results; squares correspond to the case of distinguishable 
  photons and circles to the case of indistinguishable photons. Solid lines are fits 
  of measured data and dashed lines are theoretical predictions.}
  \label{graf1}
\end{figure}

\section{Results}
The aim of the experiment is to show how the visibility of the
signal photon is affected by a distinguishable and
indistinguishable ``noise'' photon after the ``noise
subtraction''. We measured coincidence rate $C$ between detectors
D$_1$ and D$_3$. Intensity transmissivity of VRC$_2$ was adjusted
so that the visibility of coincidence rate $C$ was maximal, i.e.,
$T_{R}=1/2$ for distinguishable photons and $T_{R}=3/4$ for
indistinguishable photons. The visibility of $C$ was measured for
different values of the transmissivity $T$. $T = 100$\% represents 
no added noise case, $T = 0$\% means that the signal photon can
not pass through the lower arm of the MZ interferometer. These two
limit cases could not be measured, because the coincidence rate
$C$ vanishes.  

Figure~\ref{graf1} shows visibilities of the coincidence-rate interference 
patterns. 
Each interference-fringe measurement, consisting of 41 phase-steps, was
repeated five times. Coincidence-rate measurement for each phase step takes
typically 3~s. After each three-second measurement period the phase was
actively stabilized. 
The results displayed in Fig.~\ref{graf1} 
support the theoretical prediction that visibility $V$ does not depend on $T$.
%
Obtained mean value of visibility is $67.4\pm1.1\%$ for distinguishable 
noise photons (the theoretical value is $1/\sqrt{2}\approx70.7\%$) and 
$92.6\pm1.7\%$ for indistinguishable noise photons (the theoretical 
value is 100$\%$). 
Shown error bars represent statistical errors. 
Systematic shifts of the values are due to experimental imperfections.
It should be noted that the measurement with 
distinguishable noise photons is more robust. 
In the case of indistinguishable photons the visibility is 
very sensitive to fluctuations of the time overlap of the two photons.
Due to this fact, the visibility measured with distinguishable
photons lies closer to the theoretical limit.

\section{Conclusions}
We have observed how the noise represented by an additional
distinguishable particle can degrade interference. 
It is known that as a consequence of decoherence events very fast sudden death of
entanglement can happen and Bell-inequality violation can disappear \cite{Almeida_2007}.
So, let us imagine now that instead of a single signal photon entering the
interferometer through the input beam splitter we have a photon
inside the interferometer which is a member of a pair maximally
entangled in spatial modes. So we have two maximally entangled
qubits, one of them goes through our noisy ``channel'' followed
by the ``noise subtraction'' and finally it is measured in the
basis consisting of two orthogonal equatorial states. Pairs of
maximally entangled qubits can be used to test the exclusivity
of quantum mechanics. If they are measured locally in proper
bases (which can be fully constructed from the equatorial
states) the results violate Bell (or CHSH) inequalities
\cite{CHSH}. However, once one of the qubits is sent through
our channel and once the detectors are not able to distinguish
between modes $B$ and $B'$ no violation of the Bell inequality
is observed. To reveal the violation, a measurement outside the
equatorial plane has to be performed. It is not surprising,
since the considered decoherence process is basis dependent. It
fully disturbs only the results of measurements in the
equatorial plane. Tittel et al. \cite{tittel} used energy-time
entangled photons to test Bell-inequality violation under the
dephasing process in an optical fiber and they have proved that
the necessary condition to observe the violation reads
$V>1/\sqrt{2}$. 
In the decoherence process described in this
paper the maximal visibility (in the case of
``distinguishable'' noise) reaches just this boundary value.

\section*{ACKNOWLEDGEMENTS}
M.D. appreciates discussions with Jarom\'{\i}r Fiur\'{a}\v{s}ek.
This work was supported by the grants of the Czech Science Foundation
202/09/0747, the Czech Ministry of Education MSM6198959213 and
LC06007, and Palacky University PrF\_2010\_020 and PrF\_2011\_015.



\begin{thebibliography}{99}
  \bibitem{Zurek_03}
    W.H. Zurek, Rev. Mod. Phys. 75, 715 (2003).
  \bibitem{Schlosshauer_04}
    M. Schlosshauer, Rev. Mod. Phys. 76, 1267 (2004).
  \bibitem{Wooters_79}
    W.K. Wootters and W.H. Zurek, Phys. Rev. D 19, 473 (1979).
  \bibitem{Zurek_82}
    W.H. Zurek, Phys. Rev. D 26, 1862 (1982).
  \bibitem{Joos_85}
    E. Joos and H.D. Zeh, Z. Phys. B 59, 223 (1985).
  \bibitem{Brune_96}
    M. Brune, E. Hagley, J. Dreyer, X. Ma\^{\i}tre, A. Maali, C. Wunderlich, J.
    M.
    Raimond, and S. Haroche, Phys. Rev. Lett. 77, 4887 (1996).
  \bibitem{Pfau_94}
    T. Pfau, S. Sp\"{a}lter, Ch. Kurtsiefer, C. R. Ekstrom, and J. Mlynek,
    Phys. Rev. Lett. 73, 1223 (1994)
  \bibitem{Chapman_95}
    M.S. Chapman, T.D. Hammond, A. Lenef, J. Schmiedmayer, R.A. Rubenstein,
    E.
    Smith, and D.E. Pritchard
    Phys. Rev. Lett. 75, 3783 (1995)
  \bibitem{Sciarrino_09}
    F. Sciarrino, E. Nagali, F. De Martini, M. Gavenda, and R. Filip, Phys.
    Rev. A 79, 060304 (2009).
\bibitem{comp}
    M.A. Nielsen and I.L. Chuang, Quantum Computation
    and Quantum Information (Cambridge University Press,
    Cambridge, U.K., 2007).
 \bibitem{Banuls_07} M.-C. Banuls, J.I. Cirac, 
  and M.M. Wolf, Phys. Rev. A, 76, 022311 (2007); M.C. Tichy, F. de Melo, M. Kus,
  F. Mintert, A. Buchleitner, arXiv:0902.1684v5 (2009) 
\bibitem{Zou_1991} 
X.Y.~Zou, L.J.~Wang, and L. Mandel, Phys. Rev. Lett. \textbf{67}, 318
(1991)
\bibitem{Scully_1991} M.O. Scully, B.G. Englert and H. Walther, Nature
\textbf{351}, 111 (1991)
\bibitem{Englert_1996}  B.G. Englert, Phys. Rev. Lett. \textbf{77}, 2154 (1996)
\bibitem{HOM}
C.K.~Hong, Z.Y.~Ou, and L.~Mandel, Phys.\ Rev.\ Lett.\
\textbf{59}, 2044 (1987).
\bibitem{tittel}
  W.~Tittel, J.~Brendel, N.~Gisin, and H.~Zbinden, Phys. Rev. A
  \textbf{59}, 4150 (1999).
\bibitem{CHSH}
  J.~F.~Clauser, M.A.~Horne, A.~Shimony and R.~A.~Holt, Phys. Rev. Lett.
  \textbf{23}, 880-884 (1969).
\bibitem{Almeida_2007}  M. P. Almeida, F. de Melo, M. Hor-Meyll, A. Salles,
S. P. Walborn, P. H. Souto Ribeiro and L. Davidovich, Science \textbf{316},
589 (2007).

\end{thebibliography}
\end{document}